\documentclass[11pt,fleqn]{book} 

\usepackage{tcolorbox}
\usepackage{wrapfig}
\usepackage{url}
\usepackage[utf8]{inputenc}
\usepackage{aas_macros}
\usepackage{pdfpages}
\usepackage{multicol}
\usepackage{setspace}
\usepackage{tcolorbox}
\usepackage{chngcntr}


\usepackage[top=2.54cm,bottom=2.54cm,left=2.54cm,right=2.54cm,headsep=10pt,letterpaper]{geometry} 
\usepackage{xcolor} 
\definecolor{ocre}{RGB}{52,177,201} 

\usepackage{avant} 
\usepackage{mathptmx} 

\usepackage{microtype} 
\usepackage[utf8]{inputenc} 
\usepackage[T1]{fontenc} 

\usepackage{titlesec} 

\usepackage{graphicx} 
\graphicspath{{Pictures/}{figures/cosmo/}} 

\usepackage{lipsum} 

\usepackage{tikz} 

\usepackage[english]{babel} 

\usepackage{enumitem} 
\setlist{nolistsep} 

\usepackage{booktabs} 

\usepackage{eso-pic} 

\usepackage{titletoc} 

\contentsmargin{0cm} 
\titlecontents{chapter}[1.25cm] 
{\addvspace{15pt}\Large\sffamily\bfseries} 
{\color{ocre!60}\contentslabel[\LARGE\thecontentslabel]{1.25cm}\color{ocre}} 
{}
{\color{ocre!60}\normalsize\sffamily\bfseries\;\titlerule*[.5pc]{.}\;\thecontentspage} 
\titlecontents{section}[1.25cm] 
{\addvspace{5pt}\sffamily\bfseries} 
{\contentslabel[\thecontentslabel]{1.25cm}} 
{}
{\sffamily\hfill\color{black}\thecontentspage} 
[]
\titlecontents{subsection}[1.25cm] 
{\addvspace{1pt}\sffamily\small} 
{\contentslabel[\thecontentslabel]{1.25cm}} 
{}
{\sffamily\;\titlerule*[.5pc]{.}\;\thecontentspage} 
[]

\titlecontents{lsection}[0em] 
{\footnotesize\sffamily} 
{}
{}
{}

\titlecontents{lsubsection}[.5em] 
{\normalfont\footnotesize\sffamily} 
{}
{}
{}

\usepackage{fancyhdr} 

\fancypagestyle{plain}{\fancyhead{}} 

\makeatletter
\renewcommand{\cleardoublepage}{
\clearpage\ifodd\c@page\else
\hbox{}
\vspace*{\fill}
\thispagestyle{empty}
\fi}

\usepackage{amsmath,amsfonts,amssymb,amsthm} 

\newtheoremstyle{ocrenumbox}
{0pt}
{0pt}
{\normalfont}
{}
{\small\bf\sffamily\color{ocre}}
{\;}
{0.25em}
{\small\sffamily\color{ocre}\thmname{#1}\nobreakspace\thmnumber{\@ifnotempty{#1}{}\@upn{#2}}
\thmnote{\nobreakspace\the\thm@notefont\sffamily\bfseries\color{black}---\nobreakspace#3.}} 

\newtheoremstyle{blacknumex}
{5pt}
{5pt}
{\normalfont}
{} 
{\small\bf\sffamily}
{\;}
{0.25em}
{\small\sffamily{\tiny\ensuremath{\blacksquare}}\nobreakspace\thmname{#1}\nobreakspace\thmnumber{\@ifnotempty{#1}{}\@upn{#2}}
\thmnote{\nobreakspace\the\thm@notefont\sffamily\bfseries---\nobreakspace#3.}}

\newtheoremstyle{blacknumbox} 
{0pt}
{0pt}
{\normalfont}
{}
{\small\bf\sffamily}
{\;}
{0.25em}
{\small\sffamily\thmname{#1}\nobreakspace\thmnumber{\@ifnotempty{#1}{}\@upn{#2}}
\thmnote{\nobreakspace\the\thm@notefont\sffamily\bfseries---\nobreakspace#3.}}

\newtheoremstyle{ocrenum}
{5pt}
{5pt}
{\normalfont}
{}
{\small\bf\sffamily\color{ocre}}
{\;}
{0.25em}
{\small\sffamily\color{ocre}\thmname{#1}\nobreakspace\thmnumber{\@ifnotempty{#1}{}\@upn{#2}}
\thmnote{\nobreakspace\the\thm@notefont\sffamily\bfseries\color{black}---\nobreakspace#3.}} 
\makeatother

\newcounter{dummy}
\numberwithin{dummy}{section}
\theoremstyle{ocrenumbox}
\newtheorem{theoremeT}[dummy]{Theorem}

\newtheorem{exerciseT}{Exercise}[chapter]
\theoremstyle{blacknumex}
\newtheorem{exampleT}{Example}[chapter]
\theoremstyle{blacknumbox}

\newtheorem{definitionT}{Definition}[section]
\newtheorem{corollaryT}[dummy]{Corollary}
\theoremstyle{ocrenum}

\RequirePackage[framemethod=default]{mdframed} 

\newmdenv[skipabove=7pt,
skipbelow=7pt,
backgroundcolor=black!5,
linecolor=ocre,
innerleftmargin=5pt,
innerrightmargin=5pt,
innertopmargin=5pt,
leftmargin=0cm,
rightmargin=0cm,
innerbottommargin=5pt]{tBox}

\newmdenv[skipabove=7pt,
skipbelow=7pt,
rightline=false,
leftline=true,
topline=false,
bottomline=false,
backgroundcolor=ocre!10,
linecolor=ocre,
innerleftmargin=5pt,
innerrightmargin=5pt,
innertopmargin=5pt,
innerbottommargin=5pt,
leftmargin=0cm,
rightmargin=0cm,
linewidth=4pt]{eBox}	

\newmdenv[skipabove=7pt,
skipbelow=7pt,
rightline=false,
leftline=true,
topline=false,
bottomline=false,
linecolor=ocre,
innerleftmargin=5pt,
innerrightmargin=5pt,
innertopmargin=0pt,
leftmargin=0cm,
rightmargin=0cm,
linewidth=4pt,
innerbottommargin=0pt]{dBox}	

\newmdenv[skipabove=7pt,
skipbelow=7pt,
rightline=false,
leftline=true,
topline=false,
bottomline=false,
linecolor=gray,
backgroundcolor=black!5,
innerleftmargin=5pt,
innerrightmargin=5pt,
innertopmargin=5pt,
leftmargin=0cm,
rightmargin=0cm,
linewidth=4pt,
innerbottommargin=5pt]{cBox}

\makeatletter
\renewcommand{\@seccntformat}[1]{\llap{\textcolor{ocre}{\csname the#1\endcsname}\hspace{1em}}}
\renewcommand{\section}{\@startsection{section}{1}{\z@}
{-2ex \@plus -1ex \@minus -.2ex}
{1ex \@plus.1ex }
{\normalfont\large\sffamily\bfseries}}
\renewcommand{\subsection}{\@startsection {subsection}{2}{\z@}
{-2ex \@plus -0.1ex \@minus -.2ex}
{0.5ex \@plus.2ex }
{\normalfont\sffamily\bfseries}}
\renewcommand{\subsubsection}{\@startsection {subsubsection}{3}{\z@}
{-2ex \@plus -0.1ex \@minus -.2ex}
{.2ex \@plus.2ex }
{\normalfont\small\sffamily\bfseries}}
\renewcommand\paragraph{\@startsection{paragraph}{4}{\z@}
{-2ex \@plus-.2ex \@minus .2ex}
{.1ex}
{\normalfont\small\sffamily\bfseries}}

\usepackage[hidelinks,backref=page,pagebackref=true,hyperindex=true,colorlinks=false,breaklinks=true,urlcolor= ocre,bookmarks=true,bookmarksopen=false,pdftitle={Title},pdfauthor={Author}]{hyperref}

\newcommand{\thechapterimage}{}
\newcommand{\chapterimage}[1]{\renewcommand{\thechapterimage}{#1}}

\def\thechapter{\arabic{chapter}}
\def\@makechapterhead#1{
\thispagestyle{empty}
{\centering \normalfont\sffamily
\ifnum \c@secnumdepth >\m@ne
\if@mainmatter
\startcontents
\begin{tikzpicture}[remember picture,overlay]
\node at (current page.north west)
{\begin{tikzpicture}[remember picture,overlay]
\node[anchor=north west,inner sep=0pt] at (0,0) {\includegraphics[width=\paperwidth]{\thechapterimage}};
\draw[anchor=west] (2.46cm,-6cm) node [rounded corners=20pt,fill=ocre!10!white,text opacity=10,draw=ocre,draw opacity=1,line width=1.5pt,fill opacity=.6,inner sep=12pt]{\LARGE\sffamily\bfseries\textcolor{black}{\thechapter. #1\strut\makebox[22cm]{}}};
\end{tikzpicture}};
\end{tikzpicture}}
\par\vspace*{130\p@}
\fi
\fi}

\def\@makeschapterhead#1{
\thispagestyle{empty}
{\centering \normalfont\sffamily
\ifnum \c@secnumdepth >\m@ne
\if@mainmatter
\begin{tikzpicture}[remember picture,overlay]
\node at (current page.north west)
{\begin{tikzpicture}[remember picture,overlay]
\node[anchor=north west,inner sep=0pt] at (0,0) {\includegraphics[width=\paperwidth]{\thechapterimage}};
\draw[anchor=west] (2.46cm,-6cm) node [rounded corners=20pt,fill=ocre!10!white,fill opacity=.6,inner sep=12pt,text opacity=1,draw=ocre,draw opacity=1,line width=1.5pt]{\LARGE\sffamily\bfseries\textcolor{black}{#1\strut\makebox[22cm]{}}};
\end{tikzpicture}};
\end{tikzpicture}}
\par\vspace*{120\p@}
\fi
\fi
}
\makeatother

\usepackage{graphicx}
\graphicspath{{./figures/}}
\usepackage{appendix}
\usepackage{latexsym,amsmath,amsfonts,amssymb,booktabs}
\usepackage[font=small]{caption}
\usepackage{slashed,upgreek,amscd,cancel,tensor,color}
\usepackage{adjustbox}
\usepackage[numbers,sort&compress,square]{natbib}
\usepackage{epsfig,latexsym}
\usepackage{url}
\numberwithin{equation}{section}
\usepackage{doi}
\usepackage{subcaption}
\usepackage{mathtools}
\usepackage{upgreek}
\usepackage{stfloats}
\usepackage{afterpage}
\usepackage{multirow}

\counterwithout{footnote}{chapter}

\begin{document}
\pagenumbering{roman}

\includepdf{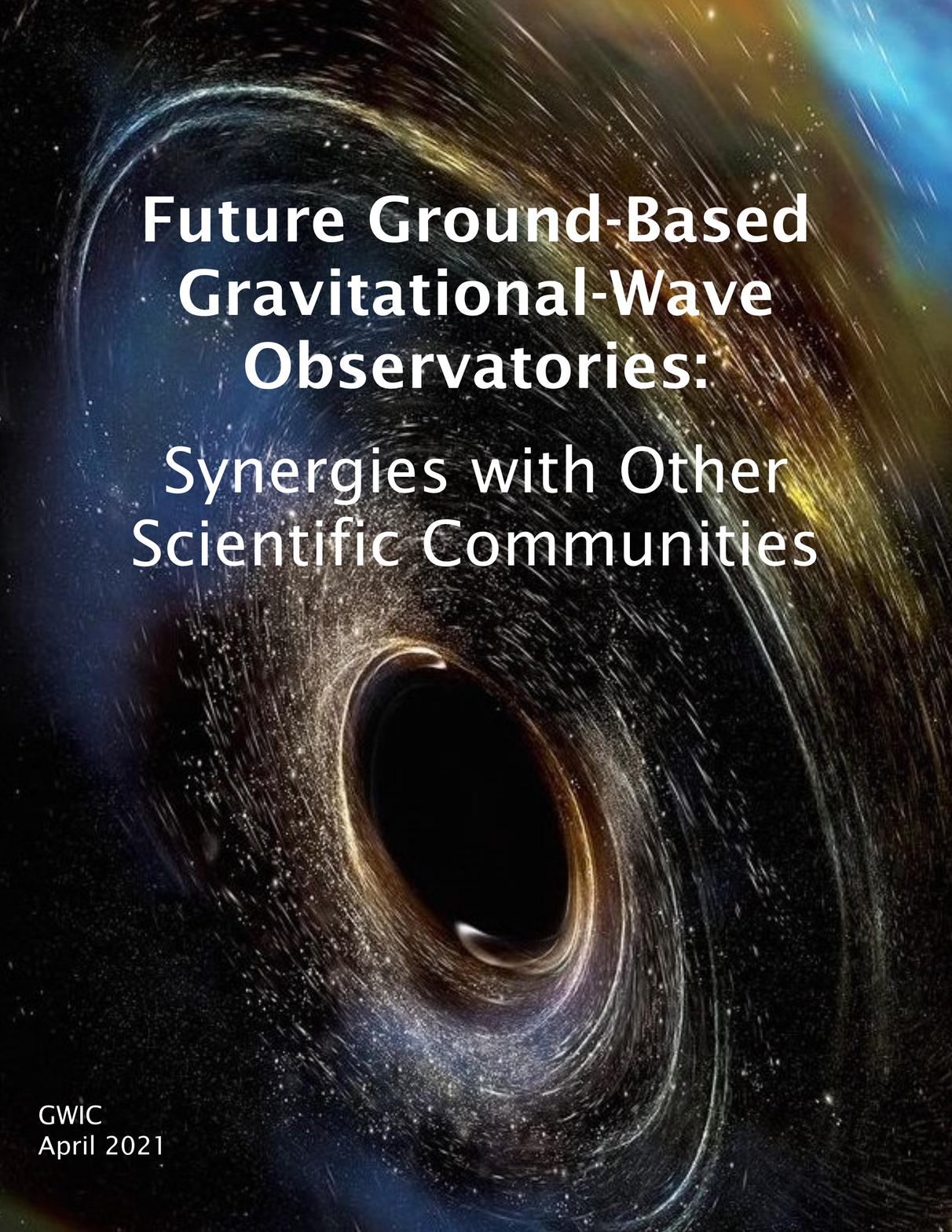}

\begingroup
\thispagestyle{empty}



\newpage
\thispagestyle{empty}

\textbf{COMMUNITY NETWORKING SUBCOMMITTEE}

Michael Punturo, INFN Perugia, Italy (Co-chair)

David Reitze, Caltech (Co-chair)

David Shoemaker, MIT, USA\\

\textbf{STEERING COMMITTEE}

Michele Punturo, INFN Perugia, Italy (Co-chair)

David Reitze, Caltech, USA (Co-chair)

Peter Couvares, Caltech, USA

Stavros Katsanevas, European Gravitational Observatory

Takaaki Kajita, University of Tokyo, Japan

Vicky Kalogera, Northwestern University, USA

Harald Lueck, AEI, Hannover, Germany

David McClelland, Australian National University, Australia

Sheila Rowan, University of Glasgow, UK

Gary Sanders, Caltech, USA

B.S. Sathyaprakash, Penn State University, USA and Cardiff University, UK

David Shoemaker, MIT, USA (Secretary)

Jo van den Brand, Nikhef, Netherlands
\vspace{9cm}

\noindent \textsc{Gravitational Wave International Committee}\\

\noindent {This document was produced by the GWIC 3G Subcommittee and the GWIC 3G Synergies with Scientific Communities Subcommittee}\\ 

\noindent \textit{Final release, April 2021}\\ 

\noindent \textit{Cover:  LIGO/Caltech/MIT/Sonoma State (Aurore Simonnet)}


\chapterimage{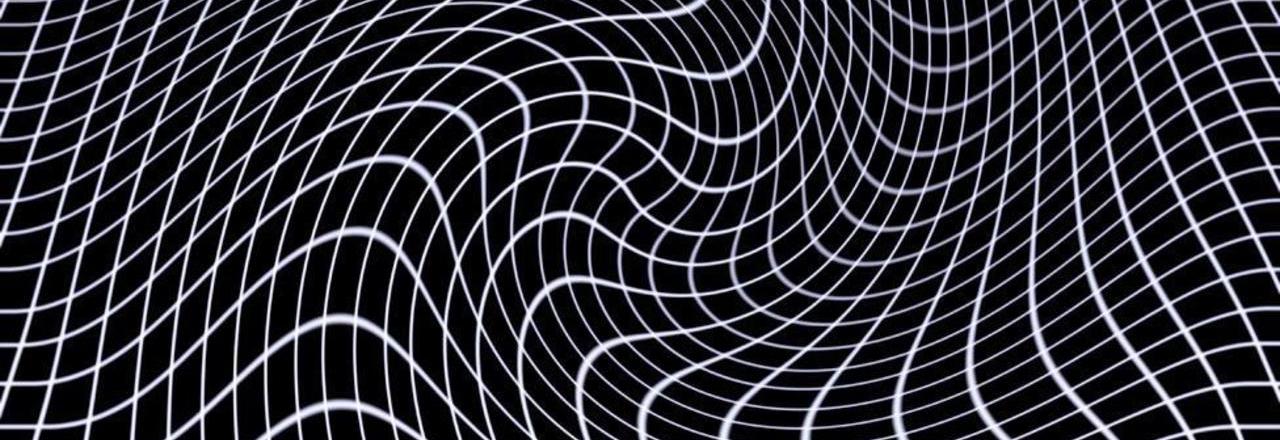} 
\pagestyle{empty} 
\tableofcontents 


\pagestyle{fancy} 
\newpage


\pagenumbering{arabic}
\chapterimage{CN_Introduction_pic.jpg} 
\chapter{Introduction}
\label{ch:introduction}

\vspace{1.0cm}
\label{sec:Intro}
\vspace{-0.05in}
{\Large\bf {P}}lanning for the development of a 3rd generation global gravitational-wave detector array is a multifaceted and complex effort that will necessarily need a high level of community input. Interfacing to extant and new stakeholders in the broader scientific constituencies is absolutely necessary to, first, keep them aware of the activities taking place in the ground-based gravitational-wave community and, second, receive input to inform and evolve the planning. The Community Networking Subcommittee within the GWIC 3G Planning Committee is charged with organizing and facilitating communications between 3rd generation projects and the relevant scientific communities. In this section, we present the approaches GWIC and gravitational-wave collaborations and projects should consider taking to engage with broader community.

As the planning for constructing the next generation of ground-based observatories matures and evolves, it is imperative that the planning process engages with the major scientific stakeholders who will utilize and benefit from the operation of those detectors.  Meaningful involvement of the broader community is an essential prerequisite in developing a ‘3G roadmap’ for the worldwide community that is well-informed by the scientific opportunities (‘science case’), a program of future detector technology development and observatory design, and an examination of potential governance models for a global array. 

The primary foci, broadly defined, are to:
\begin{itemize}
    \item raise awareness of the planning effort underway by identifying the relevant scientific constituencies and advertising the work of the GWIC 3G Committee to those constituencies. 
    \item continually and effectively communicate to those scientific constituencies the progress on planning and development as it evolves. 
    \item enable and foster the means for input and feedback from the constituencies in the planning process, and iterate with the communities as plans evolve.
\end{itemize}

In the following we outline a communication and outreach plan for engaging with and to some extent rallying the relevant communities. We emphasize that the approaches proposed here will need to continue well into the 2020s as the roadmap toward a 3rd generation detector network is finalized and executed. It will be critically important to maintain a highly visible profile for 3rd generation detectors through continued communication of the planning and progress as it occurs.

\chapterimage{CN_Constituencies_pic.jpg} 
\chapter{Constituencies \& Affiliated Communities}
\label{sec:community}

{\Large\bf {T}}he first step in developing effective approaches for community outreach is to ask the question “who are the scientific stakeholders and to what degree those stakeholders will benefit from and contribute to the scientific output of a 3rd generation global array?”.  Driven by the successful construction and operation of the Advanced LIGO and Advanced Virgo and the spectacular discoveries of merging black hole and neutron star pairs made by the LIGO Scientific and Virgo Collaborations, the community of gravitational-wave ‘users’ has grown rapidly over the past decade and will likely continue to grow into new domains now that the field is delivering scientific results.  As Figure \ref{GW-Map} illustrates, GW science is a  truly global endeavor.  

The scientific stakeholders will be the strongest advocates for making the case for building an operating the next generation of observatories.  Each interferometer node in the detector array is very likely to cost more than the entire existing generation of gravitational-wave observatories.  As such, new gravitational-wave facilities will come directly into competition with other planned large-scale scientific infrastructure projects, each with their own group of scientific proponents.  A large-scale and coherent program must be an essential element of building the case for funding a next generation gravitational-wave detector array.   

\textbf{The Ground-based Gravitational-wave Community} – This is the primary ‘user’ base for ground-based gravitational-wave detectors. Once constructed and commissioned, the 3rd generation detector network will serve as the primary instruments for the existing (and growing) ground-based gravitational-wave community. The LIGO Scientific Collaboration and Virgo Collaboration are collectively responsible for carrying out the scientific program of the LIGO and Virgo detectors. Indeed, the key to the success of the existing ground-based gravitational-wave observatories – LIGO, Virgo, GEO600 – has been the tight integration of their scientific teams and the collaborative structures that have developed and grown among the projects
over the past 25 years. 

\begin{figure}
  \includegraphics[width=13cm]{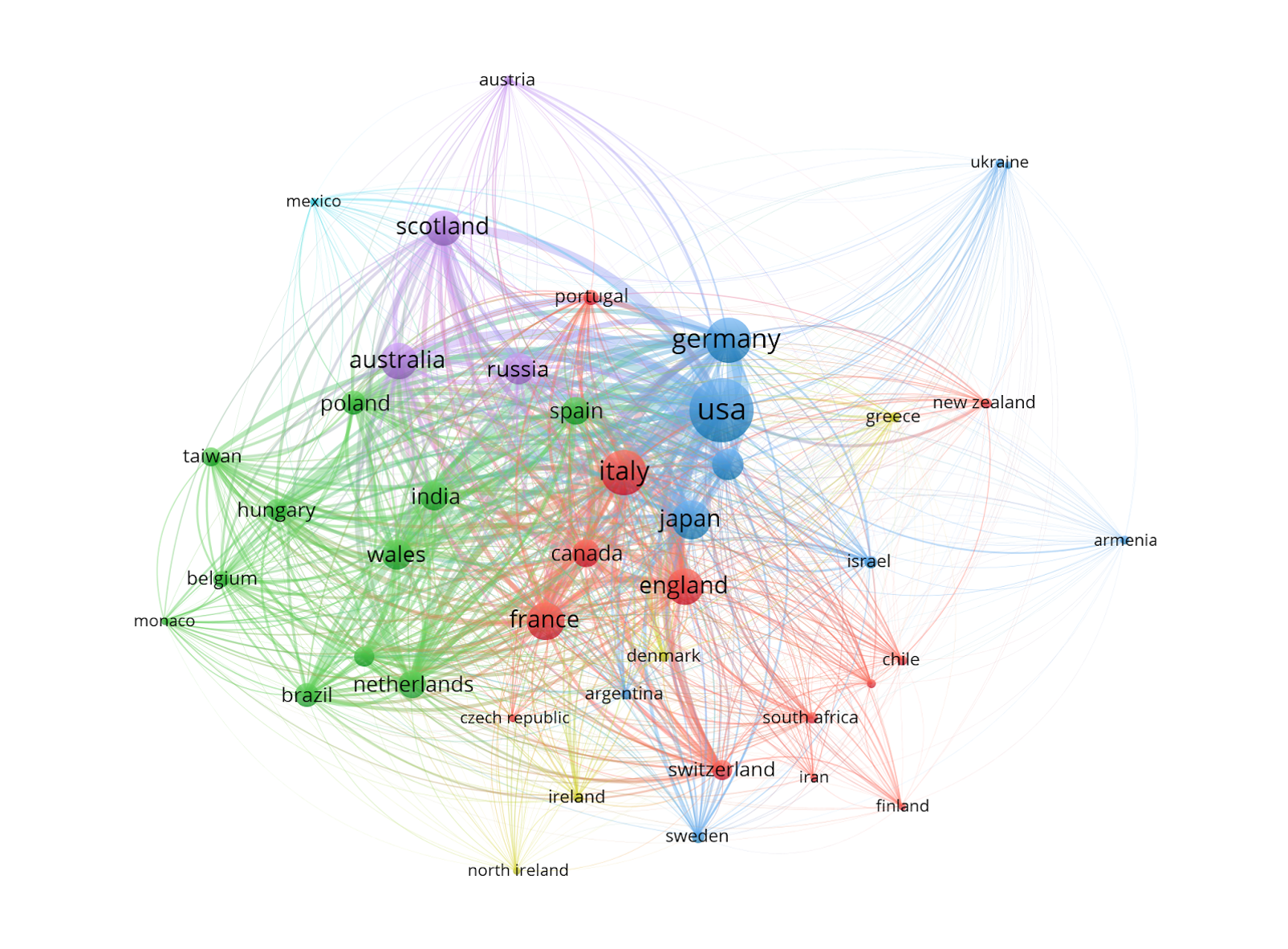}
    \centering 
    \vspace{-2mm}
    \caption{Global participation in ground-based gravitational-wave science as measured by bibliometric analysis of published journal articles, review articles, and books on GW science from 1999-2019 using the Web of Science Core Collection (WoSCC) database. (Source: Einstein Telescope Targeted Scientific Communities Study)} \label{GW-Map}
    \end{figure}

The level of engagement of the overall ground-based community with the 3G planning process is already very high. The existing collaborative framework among LIGO, Virgo, and GEO600 has been designed to grow and incorporate new 2nd generation gravitational-wave observatories as they come online. The KAGRA observatory in Japan joined the existing LIGO-VIRGO network in 2020.  LIGO-India is planned to come online toward the end of this decade.   This successful ‘collaboration of collaborations’ framework already provides the necessary structures and communication pathways for engagement and communication with the GWIC 3G Committee.   

Collectively, these existing ground-based detector collaborations are made up of more than 2000 researchers who are a natural and powerful advocacy group. In February 2018, the Einstein Telescope (ET) project underway formally established a collaboration to support the development of the 3rd generation ET detector in Europe.  The ET Collaboration will become an extremely important constituency for the 3rd generation detector network in the future.  Likewise, a collaboration is expected to develop around the US-based Cosmic Explorer (CE) effort in the near future.  

\textbf{The Space-based (LISA) and Pulsar Timing Gravitational-wave Communities} – The Laser Interferometer Space Antenna (LISA) a planned space-based detector complementary to the existing earth-based detectors, focusing gravitational sources emitting at lower frequencies than the ground-based interferometers, from $\sim$ 100 $\mu$Hz to $\sim$1 Hz.  LISA is the L3 ‘large class’ mission of the European Space Agency with NASA participation. Currently slated for launch in 2034, LISA will primarily focus on different types of gravitational-wave sources inaccessible to ground-based detectors (such as intermediate or light supermassive black hole mergers and galactic white dwarf binaries) but will have overlapping science goals with 3rd generation detectors,  thus collaborating and coordinating with the LISA community of $\sim$850 researchers will be beneficial.  LISA is a member of GWIC, thus a natural channel already exists for coordination to take place.

Gravitational-wave detection via radio telescope pulsar timing arrays (PTAs) extends the search for gravitational waves down to the nanoHertz frequency range.  Primarily focusing on supermassive black hole (SMBH) mergers, the pulsar timing array community has already produced scientifically interesting upper limits on SMBH merger rates. As with LISA, synergies exist with the PTA and ground-based communities. The pulsar timing community (NANOGrav, the European Pulsar Timing Array, and the Parkes Pulsar Timing Array Collaborations) is a member of GWIC.

\textbf{Transient Astronomy and High-energy Astrophysics Communities} - The detections of binary black hole (BBH) and binary neutron star (BNS) mergers by LIGO and Virgo over the past few years have dramatically raised awareness of communities that have heretofore not traditionally been in direct contact with the ground-based gravitational-wave community. The BBH merger GW150914 has been acknowledged by astronomers as a stunning scientific breakthrough: ``\emph{The most dramatic astronomical development of the century thus far is the detection of gravitational waves from merging black holes … during the first science run of the advanced Laser Interferometer Gravitational-Wave Observatory (LIGO).}''\footnote{\href{https://www.nap.edu/read/23560/chapter/1}{`New Worlds, New Horizons: A Midterm Assessment of the 2010 Astronomy Decadal Survey'}}  
The discovery of an optical counterpart to the BNS merger GW170817 and the subsequent massive follow up campaign by a large fraction of the transient astronomical community comprising over 2000 astronomers and astrophysics is widely considered a watershed moment in multi-messenger astronomy. 

While the electromagnetic astronomy and ground-based gravitational-wave communities have been to date very distinct from each other, they have very strong overlapping scientific interests in the arena of gravitational-wave sources containing matter.  Localization of compact gravitational-wave sources is at the heart of that joint scientific interest – a large fraction of the high energy astrophysics community covering the spectrum from gamma ray to radio astronomers are eager to receive alerts from LIGO-Virgo as well as to collaborate on selected science targets. 

The ability to provide relatively small well-defined source locations (in both sky position and distance) on rapid (minute) timescales will be one of the important pillars for the science case underlying a 3rd generation global network, and in particular for building three widely separated 3G observatories.   The transient astronomy and high-energy astrophysics communities will undoubtedly be a very important scientific constituency for future gravitational-wave observatories. 

\textbf{Cosmology, Fundamental Gravity, Dark Matter, and Dark Energy Communities} - The first detections of BBH and BNS sources have opened new gravitational-wave pathways into cosmology and precision tests of General Relativity (GR). While it is likely as the sensitivities of the existing detectors continue to improve that we will produce more precise measurements of the Hubble constant, potentially helping to resolve the tension between measurements using Type Ia supernovae standard candles and the cosmic microwave background, and put more stringent limits on GR in the dynamical strong field regime, a 3rd generation network will deliver vastly improved measurements over what can be accomplished with the existing 2nd generation network and open up new vistas for investigation. 

The limits on the difference between the speed of light and the speed of gravity determined by the near simultaneous observation of gravitational waves and gamma rays from GW170817 have already ruled out some exotic dark matter and dark energy models.  3rd generation detectors will not only improve on these results, but will potentially be able to constrain the dark energy equation of state independently of current methods using electromagnetic observations.  Similarly, the first gravitational-wave detections have produced theoretical models conjecturing that LIGO-Virgo black holes may be a component of dark matter.  While the connection between gravitational wave sources and dark matter is very preliminary and speculative at present, there is clear interest in understanding how gravitational-wave searches can shed light on the nature of dark matter. 

These aspects of gravitational-wave science touch upon many communities, including cosmology as well as numerical and analytical general relativity.   While their science programs don’t exclusively rely on the construction of a 3rd generation detector array, the obvious scientific connection to gravitational waves necessitates connecting closely to these important communities.   

\begin{figure}
  \includegraphics[width=13cm]{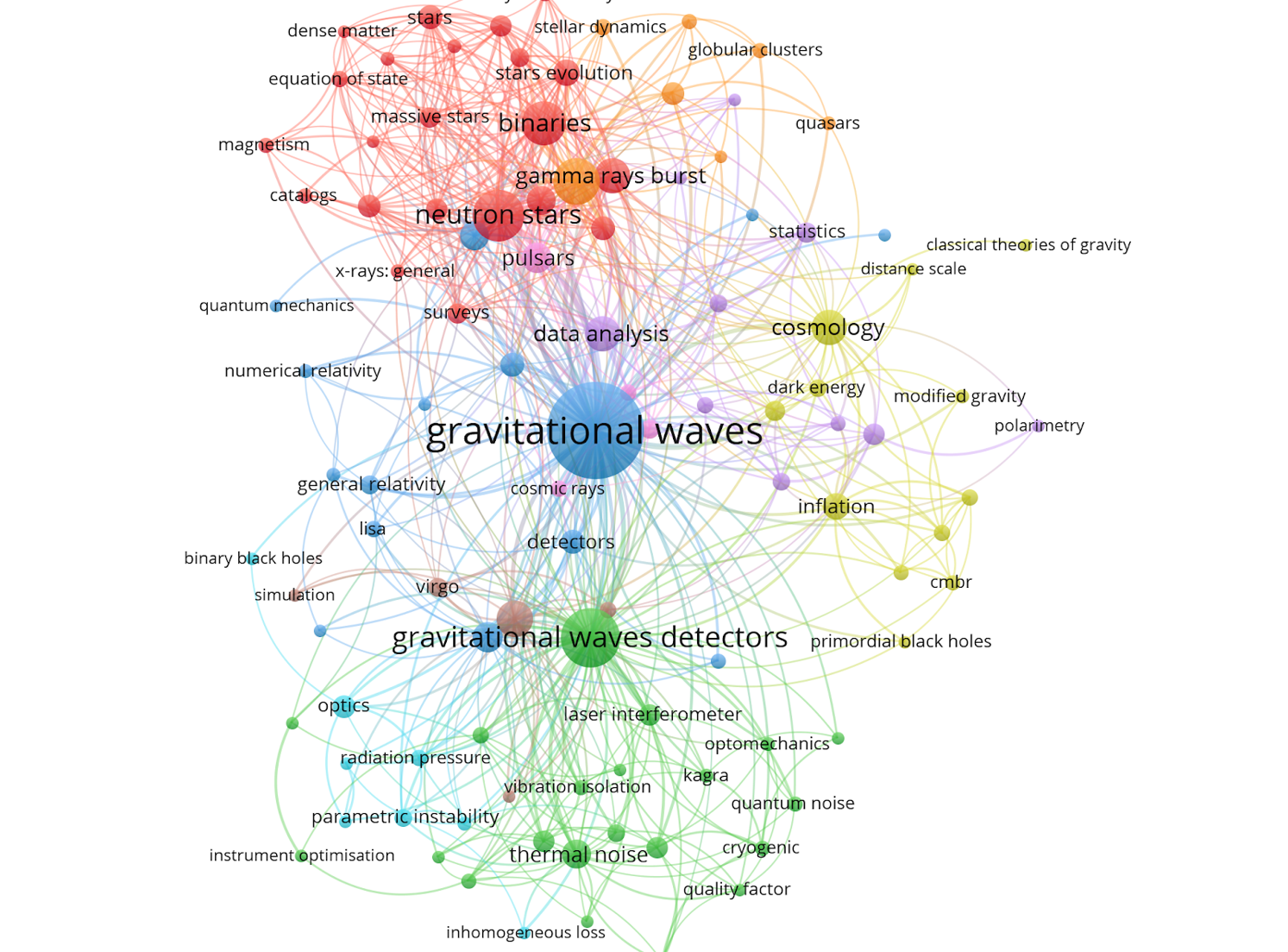}
    \centering        
    \caption{Connectivity among different disciplines related to GW science. Colors represent broad disciplines relating to astronomy and astrophysics (red, orange); gravity and general relativity (blue), high energy and cosmology (yellow), detector technologies (green). (Source: Einstein Telescope Targeted Scientific Communities Study)} \label{Communities}
\end{figure}

\textbf{Nuclear Physics Community} -The GW170817 binary neutron star merger convincingly demonstrated the common scientific synergies that exist between the nuclear physics and gravitational wave communities through new insights gained into $r$-process nucleosynthesis of heavy elements derived from observing the kilonovae produced by the GW170817 BNS collision and subsequent kilonovae dynamics. The 2015 Long Range Plan for Nuclear Science\footnote{\href{https://www.osti.gov/servlets/purl/1296778}{`2015 Long Range Plan for Nuclear Science', Chapter 4, DOE and NSF, 2015}} identifies Advanced LIGO as playing a key role in identification of r-process sites, and indeed, the groundwork for understanding the electromagnetic observations was laid in large part by the nuclear physicists and astrophysicists. The community is actively developing requirements for exascale computing to simulate nuclear processes in binary neutron star collisions.\footnote{\href{https://exascaleage.org/wp-content/uploads/sites/67/2017/05/DOE-ExascaleReport_NP_R27.pdf}{`Nuclear Physics Exascale Requirements Review'}}.   

Third generation observatories will greatly expand the ability to probe relativistic nuclear collisions at astrophysical scales. More speculatively, the discovery of exotic stars (e.g., quark, electroweak, or bosonic stars) by direct gravitational-wave observations could reveal completely new high-density states of matter.  


\textbf{High Energy Particle Physics Community} – Although the science targets of HEPP may at first glance seem somewhat distinct from the science targets of the 3G GW observatory network, scientific and technological synergies exist between the two research fields.  Common and contiguous research items can be fruitfully be explored. For example, CERN is investing growing resources into research of dark matter particle candidates like axions both within in LHC and in other detectors (e.g. the CERN Axion Solar Telescope).  The impacts of axion-like dark matter could be investigated in the coalescence of black holes and in the propagation of gravitational waves. 


More directly, HEPP has developed `Big Science' expertise and technologies that are needed by 3G detectors. Among those are large volume UHV vacuum systems, low noise cryogenics, high performance computing, and large underground infrastructure development.

These potential synergies between the needs of a 3G GW observatory network and the HEPP community led the GWIC 3G subcommittee to collaborate in the realization of a contribution for the European Strategy for Particle Physics,\footnote{\href{https://indico.cern.ch/event/765096/contributions/3295673/attachments/1785200/2906171/GW3G-ET.pdf}{`Gravitational Waves in the European Strategy for Particle Physics'}} coordinated by CERN. In addition, a Memorandum of Agreement between ET and CERN has been put into place to facilitate collaboration. 

\textbf{The Atomic, Molecular and Optics (AMO) Communities} – From a technological perspective, gravitational wave interferometers broadly overlap many research areas in atomic, molecular, and optical physics.  For example, the highly stabilized lasers used by gravitational-wave detectors find applications in atom interferometry, ultracold atom and molecule research, and precision measurement.  Gravitational-wave interferometer mirrors define the state-of-the-art in optical fabrication, coatings, and metrology.  Quantum sensing, specifically the generation and sensing of squeezed states of light, improves interferometer sensitivity beyond the shot noise and quantum radiation pressure limits and thus has natural overlap with quantum optics practitioners.  New frontiers in quantum entanglement are being explored by the gravitational-wave community as a way to make compact sub-quantum readout schemes for future interferometers.  This research has deep connections to the AMO community, as noted in the 2019 US National Academy of Sciences Report ``Manipulating Quantum Systems: An Assessment of Atomic, Molecular, and Optical Physics in the United States''.\footnote{\href{https://www.nap.edu/read/25613/chapter/1}{`Manipulating Quantum Systems: An Assessment of Atomic, Molecular, and Optical Physics in the United States'}} 

A sense of the overall span and integrated nature of GW science can be found in Figure \ref{Communities}, which displays the correlations among sub-disciplines as measured by the co-occurrence of keywords in publications relating to gravitational waves from the Web of Science Core Collection (WoSCC) database. Scaling existing technologies deployed in Advanced LIGO, Advanced Virgo, and KAGRA to the requirements imposed by the next generation of gravitational-wave detectors will further push existing limits on these technologies, thus driving a need for closer interaction and collaboration with the relevant AMO communities.    

\textbf{The General Public} - The first gravitational-wave detection captured the general public's imagination. The announcement of the gravitational waves from a binary black hole merger made the front pages of many major daily newspapers around the world on Feb 12, 2016. 
When approximately 5000 people in the United Kingdom were asked by YouGov about the discovery, 30\% of those surveyed replied that it was `exciting and important to me'.\footnote{\href{https://yougov.co.uk/topics/politics/articles-reports/2016/02/15/art-mps-and-gravitational-wawes}{YouGov Survey, Feb 12, 2016}} Intense interest by the public in gravitational-wave science continues to be driven by more recent GW detections. 

It isn't only the breathtaking astrophysics that captures the imagination.  The extreme precision with which gravitational strains are measured and the technologies developed to make those measurement also inspire fascination and awe among the public.  No one fails to be amazed when they hear that ground-based gravitational detectors measure changes in distance to better than 1/1000 the diameter of the nucleus of an atom.  Engaging the interest of the general public will provide benefits for both the public and for motivating investments in 3rd generation detector projects. The level of budget needed to build and operate 3G GW observatories will receive intense scrutiny at the highest levels of government. Broad public awareness of and support for the science afforded by 3G detectors will undoubtedly help make the case to agencies and governments that projects are worthy of support.  A robust education and public outreach 3G GW program focusing on the frontier science and technology will be a critical component in this regard.  

Finally, we point out that broad and wide-ranging nature of gravitational-wave research naturally provides excellent opportunities for enhancing inclusion and diversity in science and engineering along all axes -- race, ethnicity, gender, and geography. This should be an important consideration in education and public outreach efforts.

\clearpage
\chapterimage{CN_Engagement_pic.jpg} 

\chapter{Engaging with Constituencies \& Communities}



\vspace{8mm}
{\Large\bf {M}}eaningfully engaging with the broad array of communities identified in Section \ref{sec:community} will require a set of strategies and plans tailored to the specific constituencies. Here GWIC can play an important catalyzing role. In this sections we present some strategies and methods for interfacing to these communities. As noted in the previous section, the fundamental science targets of next generation gravitational-wave observatories span a broad set of scientific disciplines and present important opportunities for engaging the public.  A major goal of the GW community should be to ensure that those communities are \emph{aware} of the scientific capabilities and goals of future generation detectors and how those goals overlap and complement the other communities' aims.  Equally important, a coherent vision for the global third generation network should be developed and effectively communicated.

The approach to engagement will differ depending on the target audience.  Here we consider three broad audiences: scientific communities, funding agencies, and the public.

\textbf{Engaging with Constituent Scientific Disciplines} - The already high level of awareness of recent gravitational-wave discoveries makes the job of communicating the scientific potential of and opportunities afforded by 
the next generation of ground-based gravitational-wave detectors somewhat easier in principle.  Capitalizing on existing knowledge requires ensuring that the gravitational wave community continues to broadly communicate to and interact with other communities through the normal channels: participation in scientific conferences and workshops, institutional colloquia and seminars, and publishing articles (and particularly review articles) in both discipline specific and general science journals.  In addition, in Section \ref{sec:community} we presented examples of synergistic R\&D themes that can benefit both 3G GW detector development and, for example, future particle accelerators or astronomical observatories. Identifying areas of common R\&D interest and establishing collaborations with the relevant communities and laboratories will be highly beneficial.  Finally, getting advice from scientific discipline leaders and experts in other fields will prove advantageous in two ways: it is a natural informal communication channel to other disciplines and will give the 3G community valuable scientific, technical, and sociological input as the projects mature. 

\textbf{Recommendation}: GWIC and the ground-based gravitational wave projects and collaborations should proactively and continuously identify opportunities to give presentations (conferences, workshops, colloquia) and write articles (articles and popular science magazines) as a way of actively promoting the field. GWIC should take an active role in identifying appropriate conferences and reaching out to conference organizers to facilitate participation by the ground-based communities. As a first step GWIC should, working with its member projects and collaborations, develop a database and calendar of conferences and actively advertise opportunities to its membership. 

\textbf{Recommendation}: Related, GWIC should partner with the existing ground-based gravitational-wave collaborations and projects to sponsor a series of themed workshops in the US, Europe, and Asia to reach relevant constituencies as a way of both 
communicating project plans and soliciting timely scientific inputs. The existing DAWN series of workshops, and dedicated ET and CE workshops could serve in such a capacity.

\textbf{Recommendation}: Communicating the scientific opportunities and vision for a 3G network is paramount. The ground-based community should develop a standard, coherent set of presentations communicating the science opportunities and vision and emphasizing the the need for a network of 3G GW detectors.

\textbf{Recommendation}: 3G projects should seek to establish and foster collaborations with `major players' in overlapping R\&D areas and develop memoranda of agreements with laboratories where synergistic research is beneficial.

\textbf{Recommendation}: 3G projects should form advisory committees comprised of leaders from other disciplines to seek expert input on all areas relevant to future ground-based GW science.

\textbf{Engaging with National Funding Agencies} - Given the large scale and costs of building new scientific infrastructure, agencies rely on getting input from the scientific community through formal surveys, independent studies, and roadmapping exercises.  A number of such studies are described here.  Each will likely have a significant 
impact on planning for 3rd generation global array, and thus deserve special consideration.

\vspace{4mm}
\underline{The European Strategy Forum on Research Infrastructures (ESFRI) Roadmap}: ESFRI is a strategic instrument to develop the scientific integration of Europe and to strengthen its international outreach. The main scope of ESFRI is to support a coherent and strategy-led approach to policy-making on Research Infrastructures (RIs) in Europe and to facilitate multilateral initiatives leading to the better use and development of RIs. ESFRI is composed of senior science policy officials representing the Ministers responsible for Research in each of those Member States wishing to take part. A senior science policy official represents the European Commission. For many years, ESFRI has biannually updated a Roadmap for the major future Research Infrastructures in Europe; the inclusion of a project in the ESFRI Roadmap allows Member States to access specific funds from the framework programs launched by the European Commission and facilitates the financial support by the national governments. 

The deadline for the submission of ESFRI proposals is September, 9 2020 and the new roadmap is expected to be released in early 2022. A project can remain in the ESFRI roadmap for 10 years. Einstein Telescope hopes to receive the first EU funds (preparatory phase) in 2022, when the new Horizon Europe" program is expected to be operative.
 
\vspace{4mm}
\underline{The US Decadal Survey on Astronomy \& Astrophysics 2020} – Every 10 years, the US National Academy of Sciences conducts a survey of the astronomical community with the “institutional goal … to consider the past and current research of the field and provide consensus recommendations for the direction of the field over the next decade.” Decadal surveys are commissioned by NASA, NSF, and the DOE (the major US agencies funding US astronomical research) and play a critical role in influencing how to prioritize major investments in new facilities and missions in the coming decade. The Astro2020 Decadal Survey\footnote{\href{https://www.nationalacademies.org/our-work/decadal-survey-on-astronomy-and-astrophysics-2020-astro2020}{Decadal Survey on Astronomy and Astrophysics 2020 Web Site}} process began in June 2019, with the steering committee expected to issue a report sometime in mid-2021. The Astro2020 Scope\footnote{\href{https://www.nationalacademies.org/_cache_939f/content/ssb_190177-4885770000222796.pdf}{Astro2020 Scope Document}} specifically asks the committee to comment on gravitational-wave observations in the broader context of astronomy.  

This should be viewed as a two-fold opportunity.  First, the submission of science white papers is the appropriate way to communicate to the astronomy community how the ground-based gravitational-wave community is planning for the 3rd generation of gravitational-wave observatories by presenting a series of white papers highlighting the compelling science opportunities that a 3G network brings. The GWIC 3G subcommittee commissioned a series of five science white papers detailing the ‘scientific discovery space’ accessible with a 3rd generation network.  Second, the submission of a white paper presenting the technology development roadmap for interferometers (as outlined in the R\&D section of this report) in the ‘activities, projects, or state of the profession’ phase of Astro2020 is an important prerequisite to support a positive recommendation from the Astro2020 Survey on funding a technology development program, an essential step along the way to securing the needed funding for supporting Cosmic Explorer construction. An ‘Activities, Projects, or State of the Profession Consideration’ white paper\footnote{\href{https://arxiv.org/abs/1907.04833}{Cosmic Explorer: The U.S. Contribution to Gravitational-Wave Astronomy beyond LIGO}} has been submitted for consideration by the Astro2020 panel. 

\vspace{4mm}
\underline{European Strategy for Particle Physics} - In addition to the ESFRI Roadmap and Astro2020, the European Strategy for Particle Physics (ESPP) is a roadmapping exercise for particle physics carried out by CERN. Although ESPP is addressed to the high energy physics research sector, there are clear opportunities for synergistic GW research. It is important to develop the interaction mechanism with the HEPP community in order to exploit the possible synergies. To that end, the Einstein Telescope Steering Committee submitted a white paper to the ESPP highlighting the emergence of gravitational-wave physics and the overlap with particle astrophysics themes. 

The update to the ESPP was approved by the CERN Council at its June 2020 Meeting, noting that the ``groundbreaking discovery of gravitational waves has occurred since the last Strategy update, and this has contributed to burgeoning multi-messenger observations of the universe. Synergies between particle and astroparticle physics should be strengthened through scientific exchanges and technological cooperation in areas of common interest and mutual benefit."\footnote{\href{https://home.cern/sites/home.web.cern.ch/files/2020-06/2020\%20Update\%20European\%20Strategy.pdf}{2020 Update of the European Strategy for Particle Physics, European Strategy Group}}

\vspace{4mm}
\underline{`Snowmass2021' High Energy Physics Community Survey} - Analogous to the ESPP process in Europe, the US high energy physics community conducts an in-depth survey (named SnowMass) to survey the community and define the most important questions for the field and to identify the most promising opportunities to address these questions in a global context. Community input collected by SnowMass informs the Particle Physics Project Prioritization Panel (P5) which provides the long-term strategy and identifies the priorities for U.S. investments by the US Department of Energy (DOE). SnowMass2021,\footnote{\href{https://snowmass21.org/start}{Snowmass2021 website}} which began in the Spring of 2020, identifies several ares of interest in the `Cosmic Frontier' theme that have deep connections with gravitational-wave science.  These include ``CF5 Dark Energy and Cosmic Acceleration: Cosmic Dawn and Before" covering high-z gravitational-wave probes of dark energy,\footnote{\href{https://snowmass21.org/cosmic/de_first}{Snowmass2021 CF5 ``Dark Energy and Cosmic Acceleration: Cosmic Dawn and Before"}} and ``CF7 Cosmic Probes of Fundamental Physics" examining tests of general relativity with gravitational waves.

Participation in Snowmass2021 is critical for the ground-based gravitational-wave community for two related reasons.  First, it represents an established forum for engaging in a meaningful way with the high energy particle physics community, a key constituent for the 3G community. Second, as a gateway into P5, it is the primary way to establish the relevant aspects of gravitational-wave science -- cosmology, dark matter, dark energy, and tests of general relativity -- as part of the DOE program portfolio.  A prioritization of GW-related science in the next P5 report is a prerequisite for securing DOE funding and support for the 3G GW Observatory program. 

\vspace{4mm}
\underline{Discipline-specific National Academy Studies} - The US National Academies of Science Engineering and Medicine (NASEM) conduct `blue ribbon' discipline-specific studies in response to funding agency requests.  These rigorous and in-depth studies are typically convened to gauge whether an agency should invest in large scale scientific infrastructure and consider scientific potential and need, technical approaches and viability, and accuracy of cost and schedule estimates among other things.  The US NSF conducted such a study for the initial LIGO Project in 1995; it is very likely that Cosmic Explorer will need to have the imprimatur of an NASEM study as a prerequisite for construction funding.    

\textbf{Recommendation}:  The major roadmapping exercises underway in the US and Europe present critical opportunities for Einstein Telescope and Cosmic Explorer. GWIC should continue to facilitate the submission of white papers by relevant projects as well as play a coordinating role as these roadmapping exercises go forward.  

\textbf{Engaging with the General Public} - The direct detection of gravitational waves and the subsequent multi-messenger campaign for the binary neutron star GW170817 have spawned larger and more vibrant education and public outreach programs all over the world. These programs are natural vehicles for communicating the scientific opportunities afforded by the Einstein Telescope and Cosmic Explorer. The popularity of gravitational waves also leads naturally to opportunities for public talks at science festivals such as the USA Science and Engineering Festival\footnote{\href{https://usasciencefestival.org}{USA Science and Engineering Festival Web Site}} and the World Science Festival\footnote{\href{https://www.worldsciencefestival.com}{World Science Festival Web Site}} and also TED and TEDx talks.

\textbf{Recommendation}: Gravitational-wave projects and collaborations should develop outreach programs and materials specifically geared to informing and educating the general public about planned 3G gravitational wave observatories. These should focus on communicating, among other things, the amazing scientific potential of 3G detectors as well as the incredible scale and precision required. GWIC should take active steps to raise awareness of the 3G projects and science case among the formal GW-themed EPO programs, encouraging and helping EPO programs to develop materials and `story lines' for 3G science. 

\textbf{Recommendation}:  Members of the ground-based collaborations should actively seek opportunities to convey the many exciting aspects of next generation gravitational wave science (public talks, publications, social media platforms, engagement w/ government officials) as way of maintaining public interest in the field.

Somewhat related, efforts should be made in the US to identify and cultivate philanthropic organizations and science-themed foundations that seek to direct funds toward support large scale scientific endeavors.  A number of high profile examples from the astronomy  community exist, including the construction of the Thirty Meter Telescope International Observatory supported by the Gordon and Betty Moore Foundation and the Vera C. Rubin Observatory Simonyi Survey Telescope funded by Charles Simonyi and Bill Gates.  The ability to raise private monies for 3G detectors to leverage public funds could prove advantageous in achieving the necessary level of funding for next generation observatories.  The Science Philanthropy Alliance,\footnote{\href{https://sciencephilanthropyalliance.org}{Science Philanthropy Alliance Web Site}} an organization whose mission is to increase philanthropic support for basic scientific research is one possible avenue for exploring how to approach and interact with foundations and philanthropies.

\textbf{Recommendation}:  The Cosmic Explorer Project should explore opportunities avenues for seeking private funding.

\clearpage

\chapterimage{CN_Summary_pic.jpg} 
\chapter{Summary of Recommendations}


\begin{itemize}

\item  \textbf{Recommendation}: GWIC and the ground-based gravitational wave projects and collaborations should proactively and continuously identify opportunities to give presentations (conferences, workshops, colloquia) and write articles (articles and popular science magazines) as a way of actively promoting the field. GWIC should take an active role in identifying appropriate conferences and reaching out to conference organizers to facilitate participation by the ground-based communities. As a first step GWIC should, working with its member projects and collaborations, develop a database and calendar of conferences and actively advertise opportunities to its membership. 

\item \textbf{Recommendation}: Related, GWIC should partner with the existing ground-based gravitational-wave collaborations and projects to sponsor a series of themed workshops in the US, Europe, and Asia to reach relevant constituencies as a way of both communicating project plans and soliciting timely scientific inputs. The existing DAWN series of workshops, and dedicated ET and CE workshops could serve in such a capacity.

\item \textbf{Recommendation}: Communicating the scientific opportunities and vision for a 3G network is paramount. The ground-based community should develop a standard, coherent set of presentations communicating the science opportunities and vision and emphasizing the the need for a network of 3G GW detectors.

\item \textbf{Recommendation}: 3G projects should seek to establish and foster collaborations with `major players' in overlapping R\&D areas and develop memoranda of agreements with laboratories where synergistic research is beneficial.

\item \textbf{Recommendation}: 3G projects should form advisory committees comprised of leaders from other disciplines to seek expert input on all areas relevant to future ground-based GW science.

\item \textbf{Recommendation}:  The major roadmapping exercises underway in the US and Europe present critical opportunities for Einstein Telescope and Cosmic Explorer. GWIC should continue to facilitate the submission of white papers by relevant projects as well as play a coordinating role as these roadmapping exercises go forward. 

\item \textbf{Recommendation}: Gravitational-wave projects and collaborations should develop outreach programs and materials specifically geared to informing and educating the general public about planned 3G gravitational wave observatories. These should focus on communicating, among other things, the amazing scientific potential of 3G detectors as well as the incredible scale and precision required. GWIC should take active steps to raise awareness of the 3G projects and science case among the formal GW-themed EPO programs, encouraging and helping EPO programs to develop materials and `story lines' for 3G science. 

\item \textbf{Recommendation}:  Members of the ground-based collaborations should actively seek opportunities to give public talks on next generation gravitational wave science as way of maintaining public interest in the field. 

\item \textbf{Recommendation}:  The Cosmic Explorer Project should explore opportunities avenues for seeking private funding. 

\end{itemize}


\clearpage

\clearpage

\end{document}